\def\kms{\mbox{\,km\,s$^{-1}$}}
\def\etal{{\it et al.}~}
\documentstyle[11pt,paspconf,epsf]{article}

\begin{document}

\title{An HI Survey of LSB galaxies selected from the APM Survey}
\author{St\'ephanie C\^ot\'e}
\affil{Dominion Astrophysical Observatory, Herzberg Institute of Astrophysics,
National Research Council of Canada, 5071 W. Saanich Rd., Victoria, BC, 
V8X 4M6, Canada}
\author{Tom Broadhurst} 
\affil{Astronomy Department, University of California, Berkeley, CA94720, USA}
\author{Jon Loveday}
\affil{Astronomy \& Astrophysics Department, University of Chicago, 
5640 S Ellis Ave, Chicago, IL60637, USA}
\author{Shannon Kolind}
\affil{Department of Physics \& Astronomy, University of Victoria, Victoria BC, V8X 4M6, Canada}
\begin{abstract}
We present preliminary results of a neutral hydrogen (HI) redshift survey to 
find low surface brightness (LSB) galaxies in the very nearby universe. Our 
sample
consists of all galaxies in the APM catalog (Maddox \etal 1990) with a mean 
surface brightness of 
$\mu \geq 24$ mag/arcsec$^2$, down to a magnitude limit of $b_j \leq 17$.
With the Parkes 64m radiotelescope 35 objects were detected at $v < 4300$ \kms.
The resulting luminosity function, HI mass function, and for the first time
 field mass function are presented. It is found that LSBs make a negligible 
contribution to the overall integrated luminosity, HI mass, and total mass
contained in galaxies.
\end{abstract}


\section{Introduction}

Redshift surveys of the general field galaxy population have revealed an
interesting class of optically low-surface-brightness (LSB) galaxy which refuse
to yield a redshift by optical means, and which may represent however a 
significant population. Studies of nearby groups of galaxies have unveiled
a large population of neutral hydrogen (HI) rich dwarfs at low redshifts
(C\^ot\'e \etal 1997). It is natural to suppose that more of these
gas-rich dwarfs could be lurking in the field neighborhood, and could thus
be relatively easily amenable to redshift detection in the radio.
The following survey was designed to explore this possibility. Here we present
preliminary results.

Our sample was extracted from the 'APM Galaxy Survey' catalog of Maddox 
\etal (1990), which contains about 2 million galaxies spread over 4300
square degrees covering the southern galactic cap.  
Because images were identified
by the APM as connected groups of pixels (16 pixels minimum) with densities
higher than a set threshold above the local sky, the resulting catalog
 is effectively diameter-limited, to $A\simeq 4$ arcsec$^2$, with a limiting
surface brightness $\mu \simeq 24.5$ mag/arcsec$^2$. 
To generate a well-defined sample of LSB field galaxies we retained {\bf all} 
galaxies satisfying the following criteria:

$\bullet $  mean surface brightness of $\mu \geq 24$ mag/arcsec$^2$

$\bullet $ magnitude of $b_j \leq 17$

\noindent This produced a candidate list of 88 galaxies.

\section{Observations}

All our candidates were observed in HI at the 64m Parkes 
Radiotelescope, using a 32 MHz bandwidth, which resulted in a channel 
separation of 6 \kms ~for a (usable) velocity range of -300 to 4300 \kms.
This negative velocities coverage was necessary to ensure that 
unknown Local Group objects would be recovered. Our 3$\sigma $ detection limit 
is $M_{HI}=1.5\times 10^8 M_{\odot}$ (at 4300 \kms using H$_o$=75 km/s/Mpc).
This yielded 35 redshifts.

These nearby galaxies were then observed at the CTIO 1.5m with a Tek1024 CCD in
B and R, with typical exposure times of 20 minutes. Surface photometry was
performed, and luminosity profiles (down to $\mu _B \sim 26.5 $mag/arcsec$^2$) 
were fitted to obtain the structural parameters, and integrated to get 
 isophotal magnitudes, more reliable
 than the APM magnitudes.

This survey is therefore very much in the same flavor as the large APM survey
of Impey \etal (1996) who 
obtained 332 redshifts in HI at Arecibo 
and in H$\alpha$. Their sample was not magnitude-limited though and their 
diameter limit was higher ($A\simeq 104$ arcsec$^2$), which means that their
sample is more susceptible to observational selection bias, that
needed to be corrected for (Sprayberry \etal 1996, 1997). By 
restricting ourselves to a relatively bright magnitude limit we avoided being
affected by these biases (see Figure 1).
Also our HI spectra have better resolution and sensitivity since our intended
targets were nearby dwarfs. 

\begin{figure}[h]
\epsfxsize=8.5cm
\epsfysize=7cm
\centerline{\epsffile{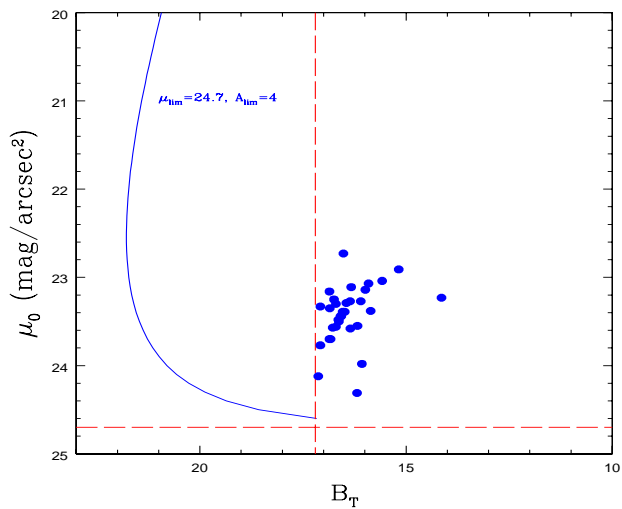}}
\caption[1]{\label{f1}
The curve shows the envelope of a selection function defined by a limiting
isophote $\mu _{lim}$=24.7 and limiting angular size of 2.25". Because of
the bright magnitude limit of our sample (B$\simeq b_j+0.2=17.2$) and the small
limiting size  we avoid being affected by the usual selection bias.}
\end{figure}

\section{Results}
The absolute magnitudes of the objects recovered in this survey range from
$M_B=$-18.5 down to -10.9, with disk scalelengths from 4 kpc to 0.15 kpc.
The average central surface brightness is 23.4 mag/arcsec$^2$, but there is 
a wide range of central surface brightnesses at a given absolute 
magnitude. The median colour of our sample is $\langle B-R\rangle $=0.93, which is 
essentially the same as that expected from High surface brightness
galaxies: de Blok \etal (1995) have analysed a sample of galaxies
extracted from the ESO-LV catalog (to use as a comparison sample
 for their LSBs) and found $\langle B-R\rangle =0.92$ for the HSBs. A wide range of colours
is recovered for our nearby galaxies, from $B-R$=0.55 to 1.22, but again 
this is typical of the range found in normal late-type galaxies. No
correlation is seen between the surface brightnesses and the colours 
  which would not be the case 
if LSBs were the faded remnants of HSBs (see the discussion in O'Neil
\etal 1997). The disk scalelengths in B and R agree within 20\% for
almost all the objects; combined with the fact that no correlation between
colour and inclination is seen, and that only one object has been detected
by IRAS, this all suggests that these LSBs have a relatively low dust
content.

 Figure 2 shows the distribution of morphological types in our sample. As
expected the majority of them are irregulars. While spirals have more 
HI than irregulars (in terms of total gas mass in $M_{\odot}$), proportionally
more irregulars are LSB galaxies, and with our HI detection limit we can recover
 'typical' irregulars down to $M=-15$, according to the magnitude-HI mass 
relations derived by Tully (1988) and Rao \& Briggs (1993). One elliptical
is also found, since following these same relations from Wardle \& Knapp (1986)
and Rao \& Briggs (1993), we should be HI sensitive to them down to a magnitude
of $M_B=-18$. The shaded histogram in the Figure shows the numbers of barred
galaxies amongst them. Contrary to previous claims the same proportion of
barred objects is seen amongst LSBs (about 34\%) compared to normal galaxies,
which are 30\% barred (van den Bergh, 1998) (see also Knezek, this 
volume).

\begin{figure}[h]
\epsfxsize=8.5cm
\epsfysize=7cm
\centerline{\epsffile{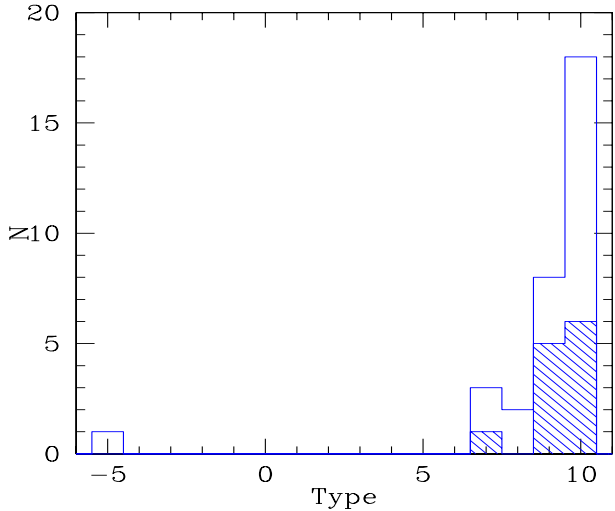}}
\caption[2]{\label{f2}
Distribution of morphological types. The hatched histogram shows
the fraction of barred galaxies in the sample.}
\end{figure}

Amongst our 35 detections only 9 objects had already published redshifts
(from the ESO catalog, the Southern Sky Redshift Survey Catalogue of da Costa 
\etal (1991), or C\^ot\'e
\etal (1997)). As far as very nearby dwarfs are concerned only 3 objects
were detected at $V_{\odot} < 1000$ \kms: ESO305-G2 (already detected in 
da Costa \etal 1991), ESO473-G24 (already known from C\^ot\'e \etal 
1997), and APM156-15-05 detected here for the fist time at $V_{\odot}=$
230 \kms.

\section{Implications}
$\bullet $ Luminosity function

After transforming our observed velocities to the Local Group frame (using
$v_{cor}=v+300\sin l \cos b$), the luminosity function was estimated using
Schmidt (1968) $V/V_{max}$ method (Figure 3). None of our galaxy has
$\langle V/V_{max}\rangle < 0.25$ and the average is $\langle V/V_{max}\rangle = 0.63$, meaning that
the sample is not suffering from serious incompleteness. A Schecter fit
yields $\alpha =-2.14$ and $M_* = -19.7$ (although with large uncertainties).
In Figure 3 our results (where the error bars plotted are just poissonian)
 compare well with the luminosity function obtained
by Sprayberry \etal (1997) with their larger sample of LSBs, for which
$\alpha =-1.42$ and  $M_* = -18.34$. Note that their raw values were 
considerably  boosted to correct mostly for surface brightness selection biases 
(Sprayberry \etal 1996),  assuming -like many other authors-
that scalelengths and central surface brightnesses of galaxies
 are uncorrelated, which is most likely not the case 
 (de Jong this volume). Nevertheless these corrections do not appear to be
so unreasonable because in the end their final values agree well with our 
raw ones (not corrected, since as 
stated above we selected our sample such as to minimise biases).

\begin{figure}[h]
\epsfxsize=8.5cm
\epsfysize=7cm
\centerline{\epsffile{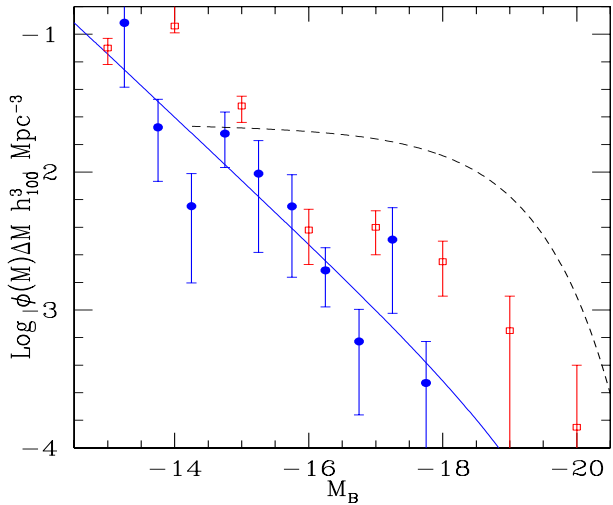}}
\caption[3]{\label{f3}
Luminosity function of our LSBs, compared to Sprayberry et al (1997) 
(squares), and Marzke et al (1994) LF for all galaxy types (dashed line)}
\end{figure}

 By integrating our luminosity function we derive 
an estimate of the total
luminosity density of LSBs of $1.6\times10^7 L_{\odot}$ Mpc$^{-3}$, while
Marzke \etal (1994) finds $11 \pm 4\times10^7 L_{\odot}$ Mpc$^{-3}$ over all
galaxy types from the CFA survey, comparable to  Loveday \etal (1992) value  
of $15 \pm 3\times10^7 L_{\odot}$ Mpc$^{-3}$. Clearly
the luminosity contribution of LSBs is only a small fraction of the total 
luminosity in the universe, despite our steep faint-end slope, because these
high number counts are for LSBs of insignificant luminosities. 

\vskip 0.2cm
$\bullet $ HI mass function

The HI spectra can yield more info than just a redshift, by integrating the
profiles one gets an estimate of the HI mass contained in LSBs. The HI 
profiles of our LSB sample have widths typical of late-type spirals and
dwarf irregulars (Schombert \etal 1992), with values for $W_{50}$ ranging from $\sim $ 20 to 203 
\kms, about
half of them exhibiting the familiar double-horned shape. As noted before
by e.g. de Blok \etal (1996) and Sprayberry \etal (1995) LSBs tend to be
more gas-rich than HSBs of the same luminosity. However in terms of total
HI mass contribution they do not seem to tilt the balance in their favour:
 Figure 4 show our HI mass function, compared to the schecter fit to the 
HI mass function of all
galaxies detected in the Arecibo Strip Survey (Zwaan \etal 1997, see also
Zwaan this volume). Galaxies with $10^9 M_{\odot}$  of HI will tend to be
normal spirals, and only a small number of them are classified as LSBs. But
down at $10^8 M_{\odot}$  of HI this is typically a late-type or an irregular,
and most of them are LSBs (de Blok \etal (1995) find an average effective
brightness of 24.25 mag/arcsec$^2$ for Sd's), which explains why our
survey basically 'catches up' with the Zwaan HI function eventually for
these low HI masses. But the majority of the HI mass in the universe is not in
LSBs but in galaxies of about $10^9 M_{\odot}$ of HI (Zwaan \etal 1997).

\begin{figure}[h]
\epsfxsize=8.5cm
\epsfysize=7cm
\centerline{\epsffile{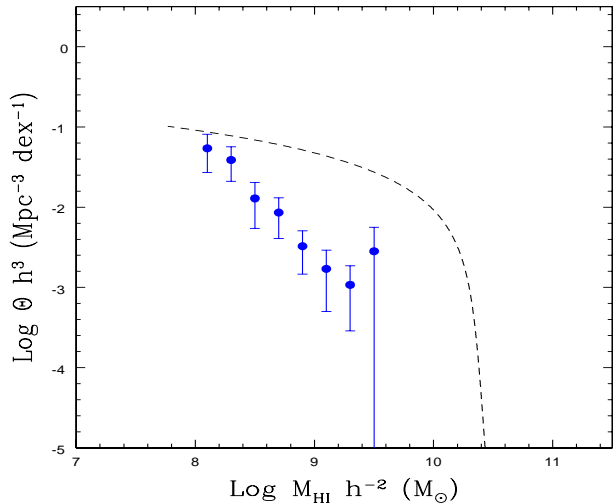}}
\caption[4]{\label{f4}
HI Mass function of our LSBs (blue dots), compared to  the schecter fit to the Arecibo
 survey data of Zwaan \etal (1997) (dashed line)}
\end{figure}

\begin{figure}[h]
\epsfxsize=8.5cm
\epsfysize=7cm
\centerline{\epsffile{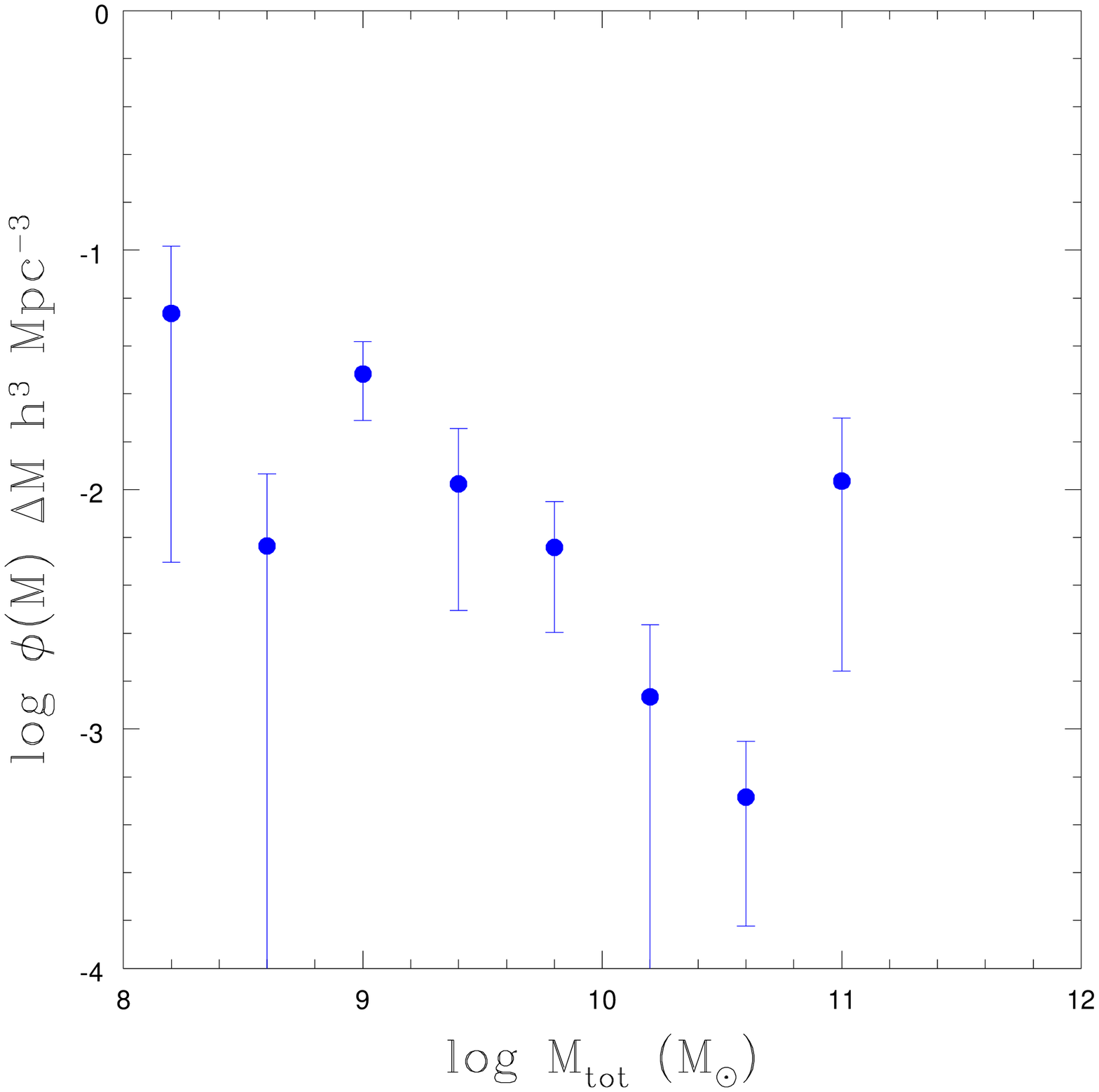}}
\caption[5]{\label{f5}
Total Mass function of our LSBs (the errorbars are from Poisson statistics)}
\end{figure}

$\bullet $ Mass function

But besides an estimate of the total HI mass of the object, an HI profile 
also reveals, from its width, something about the kinematics of the galaxy.
 From the observed $W_{20\%}$ we estimated for each galaxy its maximum 
rotation velocity $V_{max}$, by using its derived inclination from our 
photometry and
the Tully-Fouqu\'e (1985) corrections for random motions. This is still a 
reasonable thing to do for dwarfs down at $M_B=-13.25$ like in our sample, 
because they are known to be still mainly rotationally-supported at that
luminosity (see DDO154, Carignan \& Freeman 10988, also C\^ot\'e \etal 
1997, van Zee \etal 1997).
 This is the only way  (other than fully mapping them in HI with aperture
synthesis)
to estimate their $V_{max}$ because dwarfs deviate from the Tully-Fisher
relation defined by spirals (Carignan \& Freeman 1988, see also Freeman,  
this volume). With this $V_{max}$ one can then obtain an indicative 
dynamical mass with a relation of the form $M_{dyn}=R V_{max}^2/G$, where
we will use $R=7\alpha ^{-1}$, with the scalelengths derived from our
photometry, because HI rotation curves for dwarfs typically reach at least
$7\alpha ^{-1}$ (Broeils 1992). This first field mass function is presented
in Figure 6, showing the number of galaxies of a particular mass per
mass decade per Mpc$^3$.  
The rise at the faint-end is much steeper than for the mass function calculated
by Ashman \etal  (1993) who converted the luminosity function of 
Efstathiou \etal 
(1998) using the variation of the mass-to-light ratio of Salucci \etal (1992).
But it's still not steep enough for small galaxies
to dominate the mass in galaxies in the universe. With our limited  
 sample of only 35 redshifts so far this figure should be taken just as an
illustration of the interesting potential of HI surveys, compared to the 
conventional deep imaging surveys of LSB galaxies.
\acknowledgments
Many thanks to the organisers for such a pleasant and stimulating workshop.
We also thank the Parkes and CTIO TACs for all the observing time we needed.
No thanks to big corporations for polluting our radiobands which make HI
surveys like this one more and more difficult each year (see {\it Science}, 
vol282, no5386, p.34, for more details).

\end{document}